\documentclass[12pt, a4paper]{article}

\newcommand{\be}{\begin{equation}}
\newcommand{\ee}{\end{equation}}

\usepackage{graphicx}
\usepackage{amsmath}

\usepackage{a4}
\usepackage{amssymb}
\usepackage{subfigure}
\usepackage{color}
\usepackage[bf,footnotesize]{caption}

\def\xlf{\raisebox{+0.2em}{\boldmath{$\chi$}}\hspace{-0.2ex}\raisebox
{-0.2em}{L}
\hspace{-1.5ex}\raisebox{+0.14em}{F}\hspace{2mm}}

\newcommand{\tr}{\operatorname{Tr}}
\newcommand{\re}{\operatorname{Re}}

\def\desy{a}
\def\liv{b}
\def\hum{c}

\begin{document}

\begin{titlepage}
  {\vspace{-0.5cm} \normalsize
  \hfill \parbox{60mm}{LTH788\\HU-EP-08/11\\DESY 08-041\\SFB/CPP-08-23
                       }}\\[10mm]
  \begin{center}
    \begin{LARGE}
      \textbf{The  $\eta'$ meson from lattice QCD.}\\

    \end{LARGE}
  \end{center}

  \vskip 0.5cm
  \begin{figure}[h]
    \begin{center}
      \includegraphics[draft=false]{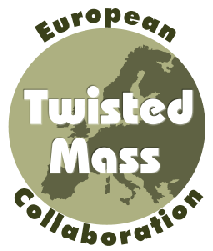}
    \end{center}
  \end{figure}

  \vspace{-0.8cm}
  \baselineskip 20pt plus 2pt minus 2pt
  \begin{center}
    \textbf{
      K.~Jansen$^{(\desy)}$,
      C.~Michael$^{(\liv)}$,
      C.~Urbach$^{(\hum)}$}\\
  \end{center}
  
  \begin{center}
    \begin{footnotesize}
      \noindent 
      $^{(\desy)}$ DESY, Zeuthen, Platanenallee 6, D-15738 Zeuthen, Germany
      \vspace{0.2cm}

      $^{(\liv)}$ Theoretical Physics Division, Dept. of Mathematical Sciences,
      \\University of Liverpool, Liverpool L69 7ZL, UK\\
      \vspace{0.2cm}

      $^{(\hum)}$ Institut f\"ur Elementarteilchenphysik, Fachbereich Physik,
\\ Humbolt Universit\"at zu Berlin, D-12489, Berlin, Germany\\      

    \end{footnotesize}
  \end{center}
  \vspace{2cm}

  \begin{abstract}
    \noindent\begin{tabular*}{1.\linewidth}{@{\extracolsep{\fill}}l}
      \hline
    \end{tabular*} 
    
    We study the flavour singlet pseudoscalar mesons  from first principles
    using lattice QCD.  With $N_f=2$ flavours of light quark, this is the
    so-called $\eta_2$  meson and we discuss the phenomenological status of
    this. Using maximally twisted-mass lattice QCD, we extract the
    mass of the $\eta_2$ meson at two values of the lattice spacing for
    lighter  quarks than previously discussed in the literature. We are
    able to estimate the mass value in the  limit of light quarks with their
    physical masses.

    \noindent\begin{tabular*}{1.\linewidth}{@{\extracolsep{\fill}}l}
      \hline
    \end{tabular*} 
  \end{abstract}

\end{titlepage}


\section{Introduction}

 There is considerable interest in understanding hadronic decays
involving  $\eta$ and $\eta'$ in the final state. The phenomenological
study of hadronic processes involving flavour  singlet pseudoscalar
mesons makes assumptions about their composition. Here we  address the
issue of the nature of the $\eta$ and $\eta'$ from QCD directly,  making
use of lattice techniques. 

 Lattice QCD directly provides a bridge between the underlying quark
description and  the non-perturbative hadrons observed in experiment. 
The amplitudes to create a given meson from the vacuum with a particular
 operator made from quark fields are measurable, an example being  the
determination of $f_{\pi}$. It also allows a quantitative study of the
disconnected quark contributions that arise in the flavour singlet
sector.   The lattice approach  provides other information such as that
obtained by varying the number of quark flavours and their masses. 

 The disconnected diagram responsible for giving the flavour-singlet 
pseudo\-scalar meson a mass (in the chiral limit) is closely related 
to the fluctuations in the topological charge. This link will be
discussed shortly in this paper and in more detail in a later
publication~\cite{etmcfollow}.

From the chiral perturbation theory description (for a review, see
ref.~\cite{Feldmann:1999uf}), one expects the mixing of $\eta$ and
$\eta'$ to be most simply  described in a quark model basis.
In the flavour singlet sector, for pseudoscalar mesons, we then have
contributions  to the mass squared matrix with quark model content $(u
\bar{u} +d \bar{d})/\sqrt{2}$  and $s \bar{s}$ (which we label as $nn$
and $ss$ respectively):
\begin{equation}
  \label{eq:qmc}
  \begin{pmatrix}
    m_{nn}^2 +2x_{nn} &  \sqrt{2} x_{ns} \\
                      & \\
    \sqrt{2} x_{ns}   &  m_{ss}^2+x_{ss}
  \end{pmatrix} \ .
\end{equation} 
 Here $m$ corresponds to the mass of the flavour non-singlet  eigenstate
and is  the contribution to the mass coming from connected fermion
diagrams while $x$ corresponds to the contribution from disconnected
fermion diagrams.  Thus $m_{nn}$ is the pion mass. Because of mixing, 
as will be discussed, $m_{ss}$ does not correspond to any specific
meson,  but its mass can be estimated assuming that the non-singlet
pseudoscalar mass-squared  is  linear in the quark mass. So $m^2_{ss}= 2
m^2_{ns}-m^2_{nn}$,  that is $ 2m_K^2-m_\pi^2$, leading to $m_{ss}=0.687
$ GeV. For a discussion from lattice results  of small corrections to
this assumption, see ref.~\cite{McNeile:2000hf}.  Chiral perturbation 
theory also gives corrections to linearity coming from loop corrections.

The mixing between the $nn$ and $ss$ flavour singlet channels must
produce the  experimental $\eta(548)$ and $\eta'(958)$. One
approximation  used historically is that of SU(3) flavour symmetry for
which  the two  physical states will be a flavour octet
$(nn-\sqrt{2}ss)$ and a flavour  singlet $(\sqrt{2}nn+ss)$. The $\eta$
is then identified as primarily  the flavour octet while the $\eta'$ is
the flavour singlet.

 Using as input $m_{nn}$ and  $m_{ss}$ and  requiring that the output
masses  ($m_{\eta}$ and $m_{\eta'}$) are correctly  reproduced, the
three mixing parameters $x$ cannot be fully determined. It is usual to
express the resulting one parameter freedom  in terms of a mixing angle,
here defined by
 \begin{equation}
  \label{eq:mixing}
  \eta =\eta_{nn} \cos \phi - \eta_{ss} \sin \phi\, , \quad 
  \eta'=\eta_{nn} \sin \phi + \eta_{ss} \cos \phi  \ .
 \end{equation}
 In our lattice study, we have two degenerate light quarks in the  sea
but no explicit strange quark. Since we do not consider partially
quenched  QCD here,  we only study the top left  corner of the mixing
matrix. The flavour singlet pseudoscalar meson (called $\eta_2$  in this
case) will then have mass-squared $m^2_{nn}+2x_{nn}$.

From the phenomenological analysis of the full mixing matrix, one can
then estimate the  mass of the $\eta_2$ meson. One such phenomenological
 analysis, motivated by lattice input (basically the magnitude of  $x$
for strange quarks), gave~\cite{McNeile:2000hf} values  $x_{nn}=0.292,\
x_{ns}=0.218, \ x_{ss}=0.13$ GeV$^2$. This assignment corresponds to  a
mixing angle $\phi$ close to $45^0$ and to a mass of the $\eta_2$ meson
of 0.776 GeV. This mass value is plausible since it is close to the
(mass-squared weighted)  average mass of the $\eta$ and $\eta'$.
Moreover, it is somewhat lighter than the   $\eta'$ meson, which has an
additional contribution to  its mass from strange quark loops
($x_{ss}$). However, here we plan to determine the $\eta_2$ meson mass 
more directly, avoiding some of the assumptions made above. 

Note that the mass of the $\eta_2$ meson comes from two components: 
$m_{nn}$ which decreases with decreasing quark mass and $x_{nn}$ which, 
in the phenomenological study, increases with decreasing quark mass. It 
is thus important to study the $\eta_2$ meson at light quark masses to
explore this. For instance, with the parameters of the phenomenological 
model described above, the $\eta_2$ meson made of light quarks of mass
equal  to the strange quark mass would have mass-squared
$m^2_{ss}+2x_{ss}$ giving  mass 0.862 GeV. Thus we can expect a rather
flat dependence of the $\eta_2$  mass on the  mass of the underlying
quarks (here a degenerate pair of valence quarks  with sea quarks of the
same nature). 

Lattice QCD is able to access flavour-singlet states, but at a cost which
 has limited these explorations in practice. Since the correlator of a 
flavour singlet meson created at $x$ and annihilated at $y$ will have
two  contributions: a connected contribution where both quark and
antiquark  propagate from $x$ to $y$ and a disconnected contribution
where there is  a quark loop at $x$ and another at $y$. This latter
contribution needs to  be measured at many values of $x$ and $y$, so it
is optimal to use a  stochastic method to achieve this. One then
evaluates the disconnected  contributions (loops) at all $x$, and from
combining them pair-wise, one can evaluate the  disconnected contribution
to the meson correlator. The stochastic method introduces noise,  which
can be reduced by appropriate variance reduction techniques. The goal is
 to reduce the noise to a level lower than the inherent noise coming
from the  underlying gauge-field variation. We shall show that the 
formalism known as twisted mass lattice QCD allows a very effective 
variance reduction method to be applied.

We present details of our lattice methods and explain the stochastic 
method used to evaluate the flavour singlet pseudoscalar meson 
correlators (both connected and disconnected). Because the signal 
from the disconnected correlator is relatively noisy, we discuss  
several strategies for reducing this error. From a combination 
of methods, we are able to present quite precise results, which enable 
us to discuss the continuum and chiral limit of the flavour 
singlet pseudoscalar mass. We compare with previous determinations 
and summarise. 

\section{Twisted mass lattice QCD action}

In the gauge sector we employ the so-called tree-level Symanzik
improved gauge action (tlSym) \cite{Weisz:1982zw}, viz.
\[
S_g = \frac{\beta}{3}\sum_x\left(  b_0\sum_{\substack{
    \mu,\nu=1\\1\leq\mu<\nu}}^4\{1-\re\tr(U^{1\times1}_{x,\mu,\nu})\}\Bigr. 
\Bigl.\ +\ 
b_1\sum_{\substack{\mu,\nu=1\\\mu\neq\nu}}^4\{1
-\re\tr(U^{1\times2}_{x,\mu,\nu})\}\right)\,  ,
\]
with the bare inverse gauge coupling $\beta$, $b_1=-1/12$ and
$b_0=1-8b_1$. The fermionic action for two flavours of maximally
twisted, mass degenerate quarks in the so called twisted
basis~\cite{Frezzotti:2000nk,Frezzotti:2003ni} reads
\begin{equation}
  \label{eq:sf}
  S_\mathrm{tm}\ =\ a^4\sum_x\left\{ \bar\chi(x)\left[ D_W[U] + m_0 +
      i\mu_q\gamma_5\tau^3\right]\chi(x)\right\}\, ,
\end{equation}
where $m_0$ is the untwisted bare quark mass, $\mu_q$ is the bare
twisted quark mass, $\tau^3$ is the third Pauli matrix acting in
flavour space and
 \[
D_W[U] = \frac{1}{2}\left[\gamma_\mu\left(\nabla_\mu +
    \nabla^*_\mu\right) -a\nabla^*_\mu\nabla_\mu \right]
 \]
is the mass-less Wilson-Dirac operator. $\nabla_\mu$ and
$\nabla_\mu^*$ are the forward and backward covariant difference
operators, respectively. 

The mass term $m_0$ is tuned to maximal twist (as described
in  ref.~\cite{Boucaud:2007uk}) at our lightest $\mu_q$
parameter. This guarantees  $\mathcal{O}(a)$
improvement~\cite{Frezzotti:2003ni}. This was shown to work excellently
in the quenched approximation~\cite{Jansen:2003ir,Jansen:2005kk}, and
there are good indications that this holds true also with dynamical
fermions~\cite{Urbach:2007rt,Dimopoulos:2007qy}.
 Moreover, 
this formalism has been found to  give a very effective way to reach
light quark
masses~\cite{Boucaud:2007uk,Blossier:2007vv,Alexandrou:2008tn}. As
such, it provides  an attractive route to evaluate the properties of
the flavour-singlet pseudoscalar mesons. 

There is one complication, however, namely that twisted mass lattice 
formalism breaks flavour and parity symmetries at finite value of the
lattice spacing $a$. These symmetries are
restored  in the continuum limit and the theory is well defined (so
providing a valid  regularisation) at finite lattice spacing. For our
present purpose, this  flavour-breaking causes the $\pi^+$ and $\pi^0$
mesons to have a mass  splitting (of order $a^2$). Moreover the $\pi^0$
has contributions from disconnected
diagrams~\cite{Jansen:2005cg,Boucaud:2007uk}.  The analysis of the
flavour-singlet  pseudoscalar meson is not significantly affected by
this and it can be  studied in the same way as was used previously in
Wilson-based lattice
formalisms~\cite{McNeile:2000hf,Allton:2004qq}. 
 In the twisted mass formalism, the $\eta_2$ meson can mix in principle
with  the neutral $a_0$ meson. This is an order $a$ mixing which could
induce  an order $a^2$ contribution to the $\eta_2$ mass observed  at
finite lattice spacing.  We expect this effect to be negligible, both
because  the $a_0$ meson is heavy and because we are working at
maximal twist. We check this by varying the lattice spacing $a$.
 For a recent review see ref.~\cite{Shindler:2007vp}.

For later convenience we introduce the following notation
 \begin{equation}
  \label{eq:Dtm}
  D_{u,d} = D_W + m_0 \pm i \mu_q \gamma_5
 \end{equation}
 for the fermion matrix of the two degenerate quark flavours (here labelled
$u$ and $d$) separately. 

The results presented in this paper are based on gauge configurations
as produced by the European Twisted Mass collaboration (ETMC). The
details of the ensembles are described in
ref.~\cite{Urbach:2007rt}. In particular, we concentrated for this
paper on the ensembles labelled $B_1, B_3, B_6$ and $C_1$ and 
$C_2$. We have compiled the details for those ensembles in
table~\ref{tab:setup}. 
The $B$-ensembles correspond to a value of the lattice spacing
of about $a\sim0.09\ \mathrm{fm}$ and the $C$-ensembles to
$a\sim0.07\ \mathrm{fm}$. The spatial lattice size is of about
$L\sim2.2\ \mathrm{fm}$ for all ensembles used here, apart from $B_6$,
which has identical parameters to $B_1$ but $L\sim2.7\ \mathrm{fm}$.

\begin{table}[t!]
  \centering
  \begin{tabular*}{0.9\textwidth}{@{\extracolsep{\fill}}lccccc}
    \hline\hline
    Ensemble & $L^3\times T$ & $\beta$ & $a\mu_q$ & $\kappa$ & $r_0/a$\\
    \hline\hline
    $B_1$ & $24^3\times 48$ & $3.9$ & $0.0040$ & $0.160856$ & $5.22(2)$\\
    $B_3$ &  & & $0.0085$ &  &\\
    $B_6$ & $32^3\times 64$ & $3.9$ & $0.0040$ & $0.160856$ &\\
    \hline
    $C_1$ & $32^3\times 64$ & $4.05$ & $0.003$ & $0.157010$ & $6.61(3)$\\
    $C_2$ & & & $0.006$ & &\\
    \hline\hline
  \end{tabular*}
  \caption{Summary of ensembles produced by the ETM collaboration used
    in this work. We
    give the lattice volume $L^3\times T$ and the values of the
    inverse coupling $\beta$, the twisted mass parameter $a\mu_q$, the
    hopping parameter $\kappa=(2am_0+8)^{-1}$ and the Sommer parameter
  $r_0/a$ in the chiral limit from ref.~\cite{Boucaud:2007uk,Urbach:2007rt}.
   The data sets cover 5000 equilibrated trajectories (10000 for $B_1$).
 }
  \label{tab:setup}
\end{table}

\section{Neutral particles in twisted mass QCD}

In quenched or partially-quenched lattice QCD, the disconnected
contribution  to the flavour singlet meson does not combine properly
with the  connected contribution to give a physical state. To avoid this
problem,  it is mandatory to study flavour singlet states in full QCD -
with  sea quarks having the same properties as valence quarks. Then the
spectrum  of flavour singlet states is well defined and can be extracted
from the  $t$-dependence of the full correlator.  Here we focus on the
case where there  are two degenerate light quarks (called $N_f=2$) which
is a consistent  theory in which to study the flavour singlet mesons.

One promising way to study  light quarks is using twisted mass QCD at
maximal twist. 
In the case of twisted mass fermions, the bilinear operator appropriate
to  create the $\eta_2$ state is $\bar{\psi} \gamma_5 \psi$ which  on
transformation into the twisted basis (used in lattice evaluation) will
become  $\bar{\chi} \tau_3 \chi$, where $\tau_3$ acts in the $(u,\ d)$
flavour space. This amounts to evaluating, in the lattice basis, the
difference of the  disconnected loop between $u$ and $d$ quarks. 
 As already reported~\cite{Michael:2007vn,Boucaud:2008xu}, this
enables a very  efficient variance reduction technique to be used to
evaluate the  disconnected diagram relevant to the $\eta_2$. 

The key observation is the following relation for the inverse $D_u$
and $D_d$:
\begin{equation}
  \label{eq:relaDuDd}
    1/D_u - 1/D_d =  -2i\mu_q (1/D_d) \gamma_5  (1/D_u)\ .
\end{equation}
 Consider now the disconnected loop $\sum X (1/D_u - 1/D_d)$ where
$X$ is some $\gamma$-matrix (here $ I$) and/or colour-matrix and  the
sum  is over space. The conventional method involves solving
$\phi_r=(1/D_u) \xi_r$ with  stochastic volume sources $\xi_r$. The
desired result is then obtained from 
 \[
 \sum X/D_u  = \sum \langle \xi^* X \phi \rangle_r\ ,
 \]
 where the average is over noise samples (labelled $r$).
However, the case mentioned above can be  evaluated efficiently using
the `one-end-trick'~\cite{Foster:1998vw,McNeile:2006bz}. Then the
required disconnected loop is given by
 \[
\sum X (1/D_u - 1/D_d) = 
-2i\mu_q \sum \langle  \phi^* X \gamma_5 \phi \rangle_r\ .
 \]
 This has  signal/noise which has a volume dependence $V/\sqrt{V^2}=1 $
which is much more  favourable than the conventional method with
signal/noise $1/\sqrt{V}$ (here $V=L^3 T$).  
 For ensemble $B_1$, we find  the zero-momentum disconnected loop at a
given time-value has a standard deviation of $\sigma=18$ for the
variation with gauge configuration and  time-slice (here called the
intrinsic variation and obtained by extrapolating to  an infinite number
of stochastic samples)    whereas the  stochastic noise on this loop has
a standard deviation of  $\sigma=87$ from 24 samples of volume source
(conventional    method) but  only  $\sigma=7.5$ from the above method
(with 12 samples). The relevant standard deviation for any analysis is
the folding  of the intrinsic variation ($\sigma=18$) with the
stochastic error. So with  a stochastic error of $\sigma=7.5$ the net
standard deviation is 19.5 whereas  with the conventional method it
would be  89. Thus there is  a significant improvement.
 So 12 inversions give the disconnected correlator from all $t$ to  all
$t'$ with no significant increase in  errors from the stochastic
evaluation.

 We also use a non-local source/sink for the meson, constructed using 
"fuzzed" links of length $6a$ as described in
refs.~\cite{Lacock:1994qx,Boucaud:2008xu},  which can be evaluated by
replacing $X$ by the corresponding fuzzed gauge links.  Thus twisted mass
lattice QCD  allows a very efficient way to evaluate flavour-singlet
pseudoscalar disconnected loops, and combining them, the required
disconnected  correlator $D(t)$.

 Evaluating the connected correlator $C(t)$ for the neutral pseudoscalar
meson as described elsewhere~\cite{Boucaud:2008xu}, then  we have the full
information $C(t)-2 D(t)$ needed to explore the spectrum in the $\eta_2$
channel. We have a $2 \times 2$  matrix of such correlators available
(local or non-local at sink/source). We could  also explore pseudoscalar
correlators  obtained by creating the $\eta_2$ state with $\bar{\psi}
\gamma_0 \gamma_5 \psi$ which  on
 twisting becomes  $\bar{\chi}  \gamma_0 \gamma_5 \chi$. This does not
allow our  very efficient variance reduction to be applied, so the
disconnected contributions  are evaluated using a hopping parameter
variance reduction~\cite{McNeile:2000xx,Boucaud:2008xu}. Even so,  they are
sufficiently noisy that we have been unable to use them to constrain the
fits or to determine the $\eta_2$ decay constant.

 We measure the auto-correlation versus trajectory number for our 
correlators. The largest auto-correlation is found for the disconnected 
contribution. To explore this more fully, we investigated the
auto-correlation  for the time-slice sum (i.e. zero momentum) of the quark
loop at a given time versus trajectory, finding  comparable 
auto-correlation times to that of the plaquette~\cite{Urbach:2007rt}.
This is not unexpected,  since the disconnected contribution is  related
to the topological charge density and thus both it and the plaquette are
sensitive to the vacuum structure encoded in the  gauge
configurations. For further discussion see ref.~\cite{Boucaud:2008xu}.
In order to cope with this measured auto-correlation, 
we block the data into sufficiently large blocks (more than  $80$
trajectories) to remove any effect on  our final error estimates. Note
that the gauge configurations we use for our measurements are usually
well separated in units of trajectories, such that even after blocking
$80$ trajectories we are left with a sufficiently large number of
(then independent) measurements.

\begin{table}[!t]
  \begin{center}

    \begin{tabular*}{.9\linewidth}{@{\extracolsep{\fill}}lccccc}
      \hline\hline
      $\beta$ & $a\mu_q$ & $L/a$ & No. & $am(\eta_2)$ & $\chi^2/\mathrm{dof}$ \\
      \hline\hline
      $3.90$ & $0.0040$ & $24$ & 888 &$0.47(8)$ & $3/(24-6)$ \\
      $3.90$ & $0.0040$ & $32$ & 184 &$0.37(6)$ & $10/(24-6)$ \\
      $3.90$ & $0.0085$ & $24$ & 245 &$0.41(7)$ & $8/(24-6)$ \\
      \hline
      $4.05$ & $0.0030$ & $32$ & 198 &$0.34(7)$ & $10/(24-6)$ \\
      $4.05$ & $0.0060$ & $32$ & 188 &$0.37(5)$ & $11/(24-6)$ \\
      \hline
    \end{tabular*}
    
  \end{center}
  \caption{Results for the flavour singlet pseudoscalar meson from  zero
    momentum correlators. The number of configurations analysed  for neutral
    correlators (connected and disconnected) is shown (No.). The fits are to
    $t$-range 3-10 with  2 states  and
    a $2\times2$ matrix of correlators (local and fuzzed at source and
    sink). The excited state mass is around $am'=1.3$ at $\beta=3.9$.
  }
  \label{table_conf}
\end{table}

\begin{figure}[t]
  \centering
  \includegraphics[width=.8\linewidth]{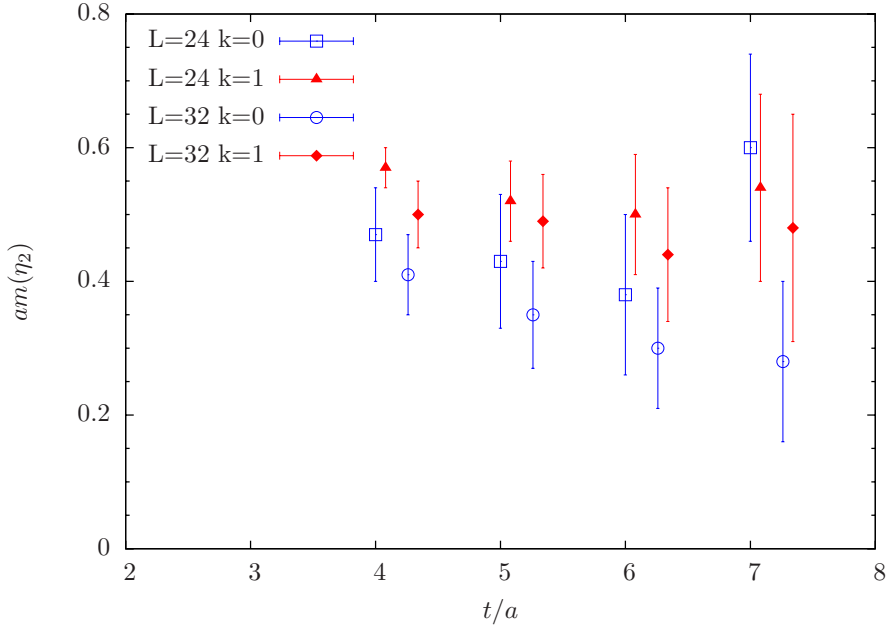} 
  
  \caption{$\eta_2$ effective mass (from a variational basis) 
    versus $t$. Results are from ensemble $B_1$ ($L=24$) and $B_6$
    ($L=32$) (both $a\mu_q=0.004$). The results from 
    momentum $k=1$ (in units of $2\pi/L$) have been plotted as mass
    values assuming $E^2=m^2+k^2$. The results are slightly displaced
    for better readability.
  }
  \label{fig:eta004}
\end{figure}

 To explore the signal, we plot the effective mass versus $t$ for
ensemble $B_1$ in fig.~\ref{fig:eta004}, where we have used the
variational basis from $t$-values $3$ and $4$ to optimise the ground state
contribution. The zero  momentum  results (open symbols) show a plateau,
although statistical errors are large, as will be  discussed later. The
results for the  two lattice spatial volumes available in this case
agree well. 

 In order to explore options to reduce the errors, we also evaluate the 
full $\eta_2$ correlator for momentum $1$ (in units of $2 \pi/L$).  Since
we evaluate the disconnected correlator for  momentum 1 in each of the 3
spatial directions, we obtain an improved estimate of $D$  (for example
with relative error  reduced to 11\% at $t=10$  compared to error of
14\% for zero-momentum). Since the symmetry classification of mesonic
states is  less sharp at non-zero momentum, the  interpretation of the
lattice spectrum is thus less straightforward.  In practice, we find
that  the small reduction in error on $D$ does not  translate into a
significant reduction in the error in  determining the underlying mass
of the $\eta_2$ state. 

 We make a fit to the combined $2\times2$ matrix of zero-momentum
correlators for  a $t$-range of 3-10 and with two states. The results
from an uncorrelated fit to blocked data are presented in
table~\ref{table_conf}. These fits are stable to  varying the $t$-range.

 Before analysing these results further,  we now discuss why the errors
are so large for the $\eta_2$ disconnected correlator, despite  the fact
that we measure all $t$ and $t'$, we use many gauge configurations   and
our stochastic errors are small. 
 As described above, we have investigated the autocorrelation (versus
trajectory) of the disconnected  contribution and find no statistically
significant evidence of any  autocorrelation beyond 40 trajectories.
We block our data into larger  blocks (typically covering $80$
trajectories) to avoid any increase in error from this source.

The origin of the  large error is that the signal for  the disconnected
part of the correlator  comes from only a small part of the total data
sample. Here we concentrate on the zero-momentum method to illustrate  this.
 We study $D_i(t'-t)$ where $i$ labels the gauge configuration and we 
explore the distribution versus $i$ at fixed $|t'-t|$.    
 For instance (for ensemble $B_1$  from $48$ $t$-values for 888 gauge
configurations) with $|t'-t|=10$,  $2.1\%$ of the gauge configurations 
contribute $26\%$ of the observed  signal (i.e. the sum over all gauge
configurations). Thus the statistical impact of the data set is smaller
than expected since parts of the data  have big fluctuations (in a
fermionic loop related to topological charge density).

Another way to illustrate this error problem is that   from the first
$444$ gauge configurations (covering $4440$ trajectories) we  find
$D/C=0.96(29)$ at $|t'-t|=16$, while for the second $444$ gauge 
configurations we find $D/C=0.47(23)$. Positivity requires $D/C<0.5$. 
So the error estimate for certain quantities from even such a big
ensemble as $5000$ trajectories can be underestimated compared to 
having $10000$ trajectories, while other quantities do not show such a
problem on the same set of gauge configurations.

So even more configurations, than we have here, would be needed to get
reliable and small errors in the case of disconnected  contributions.
This same conclusion has been obtained before, most strongly in a study 
involving staggered fermions~\cite{Gregory:2007ev}.

\section{Error reduction strategies}

Because the statistical error on the $\eta_2$ mass is still large, we 
now discuss various strategies to reduce it while retaining the same 
set of gauge configurations. Some of these strategies  do indeed reduce
the statistical error, but at the expense of introducing a systematic 
error that has to be discussed.

\subsection{Excited state removal}

 We now consider replacing the connected neutral pseudoscalar 
correlator $C$ by just the ground state contribution.  This has been 
considered previously~\cite{Neff:2001zr}. The basic idea is that the
disconnected  contributions ($D$) are only big for the lightest
flavour-singlet pseudoscalar (the $\eta_2$)  and not for excited states.
In other words, $\pi$ and $\eta_2$ are split but $\pi'$ and $\eta_2'$
are  almost degenerate. This might happen because the topological charge
fluctuations in  the vacuum, which give the flavour singlet states a
different mass, are more  strongly coupled to the ground state $\eta_2$
than to its excited states. Note that this is an assumption which 
needs to be checked by looking at the lattice results.  

 In order  to have no significant  excited state contributions in
the appropriate flavour-singlet correlator $C-2D$, one should remove
them from $C$. Then neither $C$ nor $D$ will  have excited state
contributions, and a one state fit to $C-2D$ should be possible down  to
quite low $t$-values, as indeed we shall find. 
 The extension of this argument to twisted mass lattice QCD is not trivial, 
since there are also disconnected contributions to the neutral pion itself. 
Near the continuum limit, however, the argument goes through as before.

\begin{figure}[t]
  \centering
  \includegraphics[width=.8\linewidth]{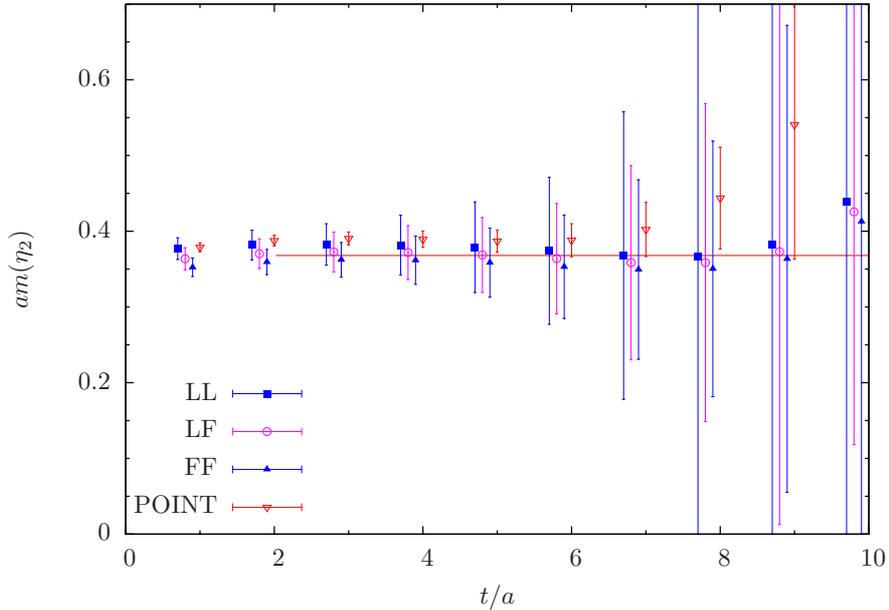} 
  \caption{Effective mass for zero-momentum $\eta_2$ correlator  for 
    ensemble $B_1$ ($a\mu_q=0.004$) taking as the connected contribution the
    ground state only. The horizontal line is the value given by a fit to the 
    correlator data in the $t$-range 2-10. Also shown 
    are results with point-to-point correlators with $r=2a$, as described 
    in the text. The results are slightly displaced for improved readability.
  }
  \label{fig:etaSESAM}
\end{figure}

In detail, we use fits of the form $c\cosh(-m(T/2-t))$ to the zero
momentum data for the connected  neutral correlator in the $t$-range
10-23 to determine the  mass $m$ and the coupling $c$. Note that this
mass $m$ corresponds to  a neutral pion only if it is interpreted as a
(partially-quenched or mixed-action) Osterwalder-Seiler
state~\cite{Jansen:2005cg}; we shall call it the `connected pion'. We 
then use our best fit parameters of $m$ and $c$ to construct the
connected pion correlator, and we assign it with an error using the
bootstrap method. This constructed connected pion correlator is finally
used together with the disconnected contribution to build the full
$\eta_2$ correlator, which we use even below $t=10$.
 As shown in fig.~\ref{fig:etaSESAM},  the  effective mass is now
essentially flat. Hence our result is consistent with dominance by a
single $\eta_2$ state down to  surprisingly small $t$-values, where the
data have small statistical errors. In contrast, fig.~\ref{fig:eta004}
shows  the effective mass obtained without this excited state removal
assumption, and  it is less flat and has larger errors, although the 
plateau value is consistent.  

 With the excited state removal assumption,  we can then make one state
fits to the resulting $2\times2$ matrix of correlators  with $t$-range
2-10 and the results are reported in table~\ref{tab:pseudo}. This shows
that we achieve a significant  reduction in the error on the $\eta_2$
mass. This reduction comes at the cost of a possible systematic error 
coming from the assumption we have made.

\begin{table}[!t]
  \begin{center}

    \begin{tabular*}{.8\linewidth}{@{\extracolsep{\fill}}lcccc}
      \hline\hline
      $\beta$ & $a\mu_q$ & $L/a$ & $am(\eta_2)$ & $\chi^2/\mathrm{dof}$ \\
      \hline\hline
      $3.90$ & $0.0040$ & $24$ & $0.37(5)$ & $0.2/(27-3)$ \\
      $3.90$ & $0.0040$ & $32$ & $0.30(4)$ & $2.4/(27-3)$ \\
      $3.90$ & $0.0085$ & $24$ & $0.36(5)$ & $3.0/(27-3)$ \\
      \hline
      $4.05$ & $0.0030$ & $32$ & $0.26(5)$ & $1.8/(27-3)$ \\
      $4.05$ & $0.0060$ & $32$ & $0.26(3)$ & $0.8/(27-3)$ \\
      \hline
    \end{tabular*}
    
  \end{center}
  \caption{Results for the flavour singlet pseudoscalar mass from
zero-momentum  correlators. The connected correlator is taken from the
fitted ground state component. The fits are to $t$-range 2-10 with  2
states  and  a $2\times2$ matrix of correlators (local and fuzzed at
source and sink).}
  \label{tab:pseudo}
\end{table}

\subsection{Point to point correlators}

 We discussed above the possibility of using correlators at smaller $t$
to reduce errors.  Another approach, which we now discuss, is to reduce
errors at all $t$-values by focussing  on a different quantity.
 The method we describe has been used recently~\cite{DeTar:2007ni} in a
study  of the disconnected contributions to the $\eta_c$ meson. 

\begin{figure}[t]
\begin{center}
        \includegraphics[width=.8\linewidth]{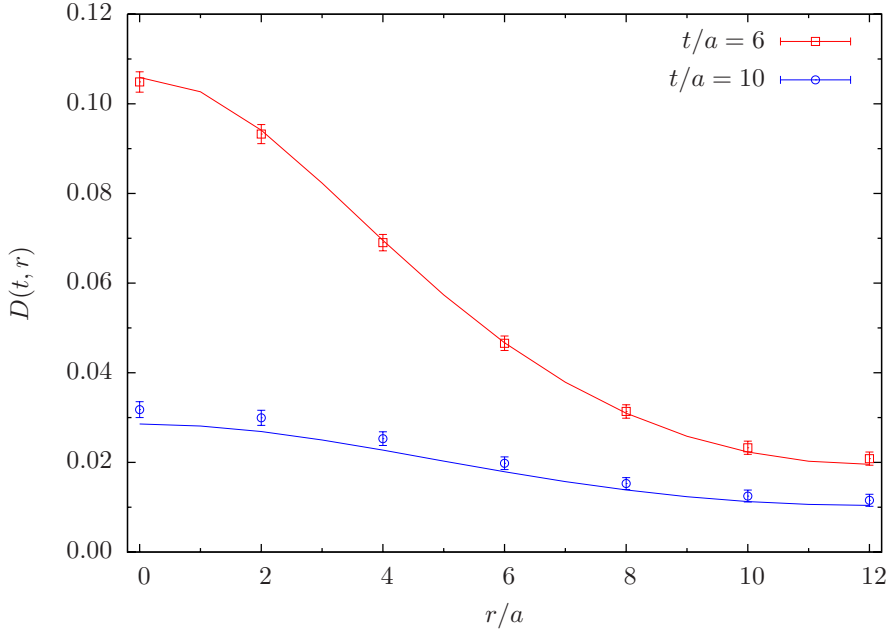} 
\end{center}
 \caption{Disconnected contribution to the $\eta_2$ correlator at 
time $t$ and spatial separation $r$ from point source and sink.
 Results are from ensemble $B_1$ ($a\mu_q=0.004$) from 319 gauge
configurations.
 The curves are from the model fit described in the text.
 }
 \label{fig:d.v.r}
\end{figure}

It is mandatory to use an all-to-all method to estimate the disconnected
loops  $d({\bf x},t)$ at all ${\bf x}$ and $t$. The disconnected
 correlator of interest reads 
 \[
D({\bf x-y}, t-t')= \langle d({\bf x},t)d({\bf y},t') \rangle ,
 \]
 assuming  that $d({\bf x},t)$ has no vacuum expectation value,  as is
the case for the flavour-singlet pseudoscalar meson because of 
the combined flavour-parity symmetry of TMQCD. 
 The approach we have considered above is to sum $D$ over the spatial
coordinates  ${\bf x}$ and ${\bf y}$. This has the advantage that the 
zero momentum correlator is studied which has the  simplest theoretical 
interpretation: as a  sum of exponentials for each state. Now $D$ is
actually peaked when $|{\bf x-y}| \approx 0$  and is small at large
$|{\bf x-y}|$. Moreover we find that the absolute error is largely 
independent of $|{\bf x-y}|$, thus the relative error is smallest
when  $|{\bf x-y}| \approx 0$.  We illustrate the dependence of $D$ on 
$r=|{\bf x-y}|$ in fig.~\ref{fig:d.v.r}. The zero momentum  contribution
is a sum over this $r$-distribution with weight approximately $r^2$ and
this enhances the larger $r$ region  which has larger relative statistical
errors. Basically the  zero-momentum sum picks up (inherent gauge-time
variation) noise by  summing over large $r$ where the signal is
unimportant but the noise is  still significant.

 For example,  we find for ensemble $B_1$ with $|t-t'|=10$ that the
relative error on $D$ when  $r=0$ (${\bf x}={\bf y}$) is  50\% of the
relative error 
on $D$ summed over both ${\bf x}$  and ${\bf y}$. Thus  the peak signal
to noise is twice as big as that summed over relative spatial 
separation.
 It might be thought that taking $r \ne 0$ would give an even smaller
relative  error, since 6 spatial directions are averaged (for spatial
separations on axis). Although we find less error reduction 
than $1/\sqrt{6}$ for $r \ne 0$, we do make use of a further small 
error reduction and we select $r=2a$  (on axis).

 We now discuss how to extract the conventional physics (i.e. mass values) 
from correlators at fixed $r$ (rather than summed over all $r$). Consider 
a correlator corresponding to a meson of mass $m$.
 In the region of $t$ where the ground state contribution dominates,  we
expect the correlator to be dependent explicitly on the meson mass when 
point operators are used at source and sink. Indeed  one way to extract
this behaviour is by evaluating the  4-dimensional lattice Fourier
transform:
\begin{equation}
  C(t,\mathbf{r})=\frac{1}{L^3 T} \
  \sum_{p_0,{\bf p}}\, \frac{c e^{-i p_0 t -i {\bf p . r}}} 
    {\hat{p}^2+m^2},
  \label{equ:4ft}
\end{equation}
 to model the data, where, to respect the periodic boundary conditions
(for a meson), $p_i=2\pi n_i/L$ and  $p_0=2 \pi n_0/T$. In
eq.~\ref{equ:4ft}, we take $\hat{p_i}=2\sin(p_i/2)$ which corresponds to
the most local lattice realisation of  the derivative in the Klein
Gordon action. Other  implementations of the derivative would shift the
mass by corrections  of order $(ma)^2$ as discussed long
ago~\cite{Michael:1982gb}.
 We have checked that this expression correctly reproduces  the
$r$-dependence for the charged pion correlator in the $t$-region  where
the ground state dominates, using the mass value  previously determined
from fitting the zero momentum correlator. Thus, though less familiar, one 
can use the expression of eq.~\ref{equ:4ft} to extract the mass value 
from the $t$-dependence at fixed $r$, in place of the usual
expression $c(e^{-mt}+e^{-m(T-t)})/(2m)$ used at zero momentum.

 To exploit this error reduction for the $\eta_2$, we combine the
disconnected  contribution (with $r=2a$ as discussed above) with the
neutral pion connected contribution also evaluated with $r=2a$. This we 
evaluate using point sources. However, it is optimal, as discussed in
the preceding section,  to use only the  ground state contribution to
the connected neutral pion  correlator, as we now discuss.

 For this  ground-state connected component, we use eq.~\ref{equ:4ft}
with as input the mass $m$ and  coupling $c$ determined from the
zero-momentum fit to the neutral (connected) pion.
 The dependence of the resulting $\eta_2$ correlator on $t$ (at fixed 
$r=2a$) can then be parametrised again by  eq.~\ref{equ:4ft}. One way to
 illustrate this, as shown in fig.~\ref{fig:etaSESAM}, is by plotting an
`effective mass' (here defined as the mass parameter that solves the
dependence  given by eq.~\ref{equ:4ft} for two adjacent $t$-values). 
 This shows that we again have a rather constant `effective mass' even
from low $t$-values, consistent with  a description by one state only.
Moreover, the errors on the `effective mass' are  smaller than from
the fixed momentum methods used above.

\begin{figure}[t]
  \centering
  \includegraphics[width=.8\linewidth]{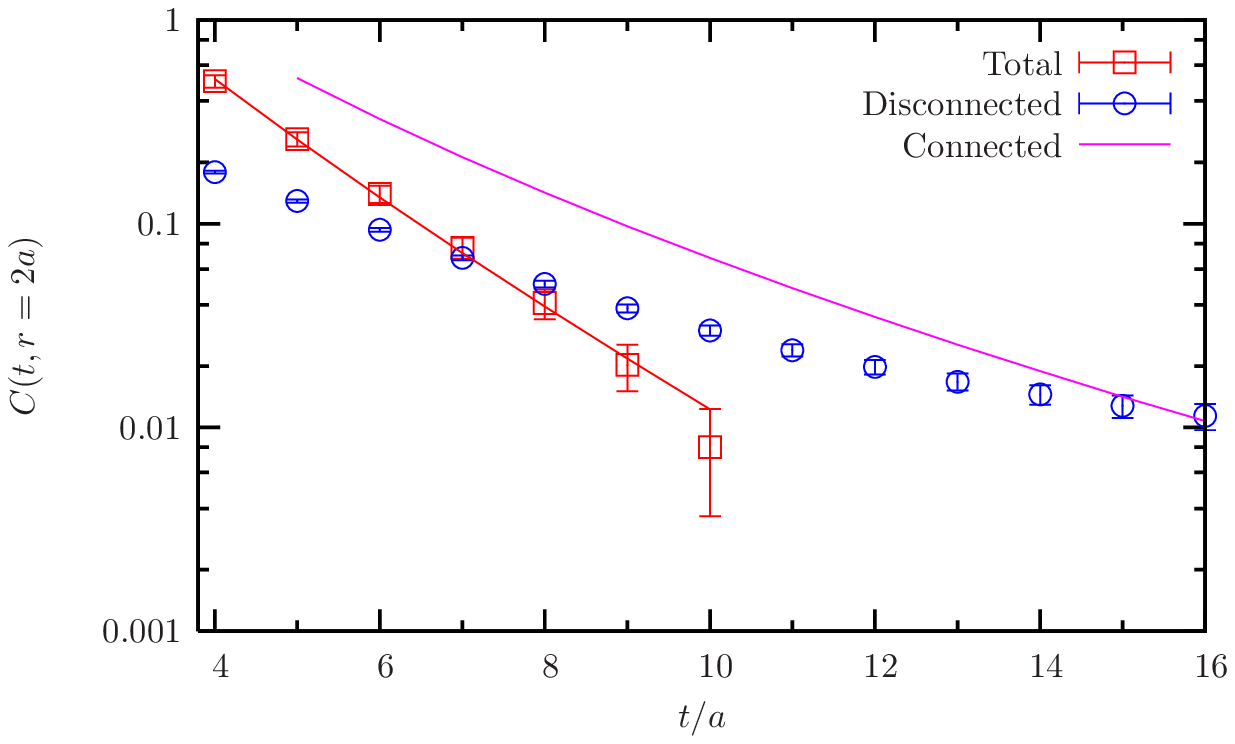} 
  \caption{Singlet pseudoscalar correlator for ensemble $B_1$ ($\mu_q=0.004$)
    with local point sink and source. The correlator is illustrated for
    spatial separation $r=2a$ versus time $t$. The upper solid line is the
    connected contribution from  the ground state neutral meson while the
    circles show the disconnected  contribution. The squares are the net
    contribution to the $\eta_2$ and the  behaviour from a one state fit to
    them is also shown. Note that the one state fit is not exponential, 
    as discussed in the text.
  }
  \label{fig:eta2_point}
\end{figure}

To determine the $\eta_2$ mass by this method,  we then  fit the
resulting combined $\eta_2$ correlator to a two parameter expression
(free mass and coupling) according to eq.~\ref{equ:4ft}. 
 Since this approach is less familiar, we illustrate in 
fig.~\ref{fig:eta2_point}  the fit versus $t$ and also in
fig.~\ref{fig:d.v.r}   the $r$-dependence of the disconnected
contribution compared to  the measured data. 
 In detail, we use fits to the zero momentum data for the connected
neutral correlator for the $t$-range 10-23 to determine the connected
pion mass $m$ and the coupling $c$, as above. From the bootstrap
ensemble of values of $m$ and $c$,  we determine, at each $t$-value, the
point-to-point connected pion correlator at fixed $r=2a$ with its
associated  error. This is then used together with the measured
point-to-point disconnected contribution  to  construct the full
$\eta_2$ correlator. The result is then fitted with one state using
eq.~\ref{equ:4ft} and a $t$-range 2-10 to obtain the $\eta_2$ mass value
 given in table~\ref{tab:point}.

\begin{table}[!t]
  \begin{center}

    \begin{tabular*}{.8\linewidth}{@{\extracolsep{\fill}}lcccc}
      \hline\hline
      $\beta$ & $a\mu_q$ & $L/a$ & $am(\eta_2)$ & $\chi^2/\mathrm{dof}$ \\
      \hline\hline
      $3.90$ & $0.0040$ & $24$ & $0.403(21)$ & $2.0/(9-2)$ \\
      $3.90$ & $0.0040$ & $32$ & $0.365(19)$ & $0.1/(9-2)$ \\
      $3.90$ & $0.0085$ & $24$ & $0.380(18)$ & $0.7/(9-2)$ \\
      \hline
      $4.05$ & $0.0030$ & $32$ & $0.302(24)$ & $0.0/(9-2)$ \\
      $4.05$ & $0.0060$ & $32$ & $0.308(18)$ & $0.0/(9-2)$ \\
      \hline
    \end{tabular*}
    
  \end{center}
  \caption{Results for the flavour singlet pseudoscalar mass using 
point-to-point correlators for the disconnected part and extracting the
connected  part from the fitted  ground state component.}
  \label{tab:point}
\end{table}

 Using  different  $t$-ranges gives the same mass value, within errors.
Compared to using the  zero-momentum approach described previously, the
statistical error on the mass is much smaller.  Here we have used local
source and sink operators  (since in this case  the  fixed $r$ behaviour
is given without extra parameters). What is especially helpful, however,
is that the resultant  data is well fitted by only one contribution for
$t>a$.

We have presented this new and powerful method for one ensemble ($B_1$)
but we  also apply it (using $r=2a$) to the other cases considered and
the results are collected  in table~\ref{tab:point}.

 This point-to-point approach provides a useful reduction in the
statistical error, though at the  potential cost of introducing a source
of systematic error in relating the quantity  measured to that actually
required. For local  meson creation and destruction operators, the
$t$-dependence at  fixed $r$ is given by the same parameters  as the
conventional zero-momentum case, up to possible non rotation-invariant
contributions of order $a^2$  which we expect to be insignificant at
large $t^2+r^2$. 
 For spatially smeared (or fuzzed) operators this  $O(4)$ invariance is
not present and we would need more parameters to  describe the
point-to-point correlators. We also find strong evidence of ground state
dominance, which enables us to use the more precise data from smaller
$t$-values. 

\begin{figure}[t]
  \centering
  \subfigure[\label{fig:etaSa}]{\includegraphics[width=.45\linewidth]{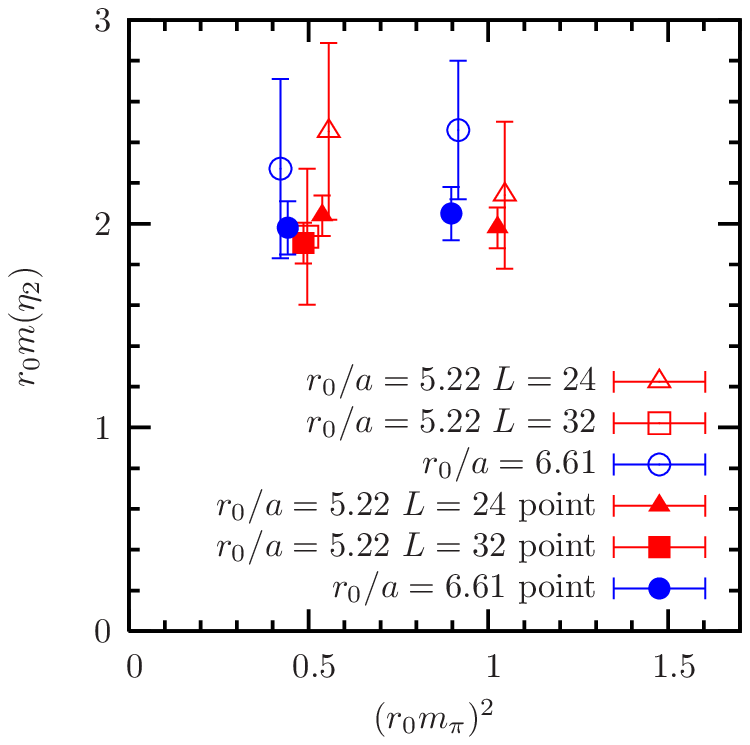}}\quad
  \subfigure[\label{fig:etaSb}]{\includegraphics[width=.455\linewidth]{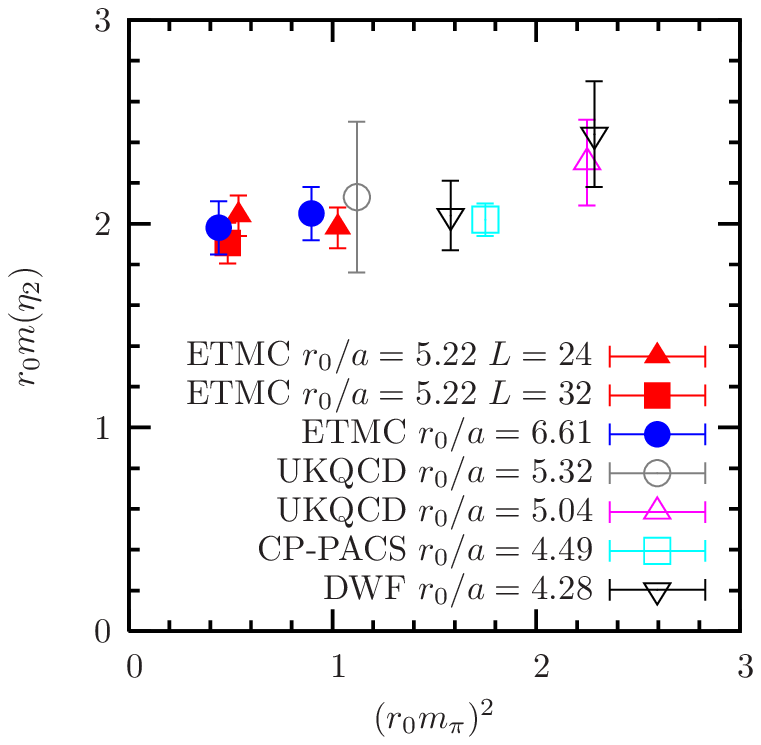}}
  
  \caption{Comparison of values for the $\eta_2$ mass versus quark mass
    in units of $r_0 \approx 0.45$  fm. Values of the lattice spacing used
    are  shown by listing $r_0/a$. Some of the points have been displaced 
    horizontally for legibility.
    The ETMC results shown in (a) are those from the 2-state fits to
    zero-momentum correlators (first method in text, open symbols)
    and from the error-reduction  obtained by  using   point-to-point correlators and 
    projecting to the ground state of the   neutral pseudoscalar connected
    contribution (filled symbols).
    In (b) we compare our results with reduced errors (filled symbols)
    to other $N_f=2$ results from other
    collaborations~\cite{Allton:2004qq,Lesk:2002gd,Hashimoto:2008xg}
    (open symbols)
    with  relatively light pions and small lattice spacing ($a < 0.1$  fm, $r_0/a
    > 4.5$). The strange quark corresponds to $(r_0 m_{\pi})^2  \approx 2.5$.   
  }
  \label{fig:etaS}
\end{figure}

\section{Summary of $\eta_2$ mass}

In order to compare the error-reduced results from different lattice
spacings, we plot them using the $r_0/a$-values given in
table~\ref{tab:setup} to create dimensionless quantities.
We see in fig.~\ref{fig:etaSa}  that the results are consistent with each
 other and are also consistent with the results presented above using 
the standard fixed-momentum analysis technique. This increases the 
statistical impact of our results.

 We also plot  previous $N_f=2$
results~\cite{Lesk:2002gd,Allton:2004qq,Hashimoto:2008xg} that are in
the quark mass and lattice spacing region we are exploring in
fig.~~\ref{fig:etaSb}. Different groups using different lattice
formulations obtain consistent results. 
 Results from our two lattice spacings are consistent with each other
and  we are able to evaluate the flavour singlet mass closer to  the
continuum limit (and in an order $a^2$ improved formalism since we work 
at maximal twist) than  hitherto.  

An extrapolation to the chiral limit of our results gives a value  of
the  $\eta_2$ mass of $r_0 m(\eta_2)=1.99(15)$ assuming a  linear
dependence. From our lattice results, we actually see no statistically
significant  evidence for any slope,  and if we assume a constant
behaviour, we would obtain  $r_0 m(\eta_2)=2.00(5)$.   
 Using $r_0=0.454(7)$fm obtained~\cite{Boucaud:2007uk} from  the ETMC
evaluation of $f_{\pi}$,  we obtain an  $\eta_2$ mass of 0.865(65) GeV 
where the error reflects that from the linear extrapolation. 
 This updates the phenomenological estimate of 0.776 GeV 
discussed above, which was in turn mainly motivated by lattice results 
at heavier quark masses than those we now have available. 
 We have additional systematic errors which are not fully under control:
from  the chiral extrapolation, the excited state removal assumption,
the continuum extrapolation and any possible volume dependence. We do
not see any sign that these  systematic errors are required to our
current statistical precision, but, as a conservative estimate, we
consider these  systematic errors to be comparable to the statistical
error. So we  quote an  $\eta_2$ mass of 0.865(65)(65) GeV.


One of the main conclusions of our study is that we find strong evidence
that  the flavour singlet mass remains finite as the quark mass is
reduced  to the chiral limit. This has implications for the topological
charge:  it implies that the topological charge susceptibility must
decrease as  $m_\pi^2$ in the chiral limit~\cite{Leutwyler:1992yt}. We
will discuss our results for topological  charge more fully
elsewhere~\cite{etmcfollow}.

\subsubsection*{Acknowledgments}

We thank Jaume Carbonell and Mariane Brinet for assistance with making 
point source propagators available for testing point to point 
propagation.  We thank all members of ETMC for useful comments, help
and a fruitful collaboration. One of the authors (CM) acknowledges 
a useful discussion with Paul Rakow.
 The  computer time for this project was made available to us by the
John von Neumann-Institute for Computing on the JUMP and  Jubl systems
in J\"ulich and apeNEXT system in Zeuthen, by UKQCD
on the QCDOC machine at Edinburgh, by INFN on the apeNEXT systems in Rome, 
by BSC on MareNostrum in Barcelona (www.bsc.es), by the NW Development 
cluster at Liverpool 
and by the Leibniz Computer centre in Munich on the Altix system.  We
thank these computer centres and their staff for all technical  advice
and help.  
 On QCDOC we have made use of Chroma~\cite{Edwards:2004sx} and
BAGEL~\cite{BAGEL} software and we thank members of UKQCD for
assistance.

This work has been supported in part by  the DFG 
Sonder\-for\-schungs\-be\-reich/ Transregio SFB/TR9-03  and the EU Integrated
Infrastructure Initiative Hadron Physics (I3HP) under contract
RII3-CT-2004-506078.  We also thank the DEISA Consortium (co-funded by
the EU, FP6 project 508830), for support within the DEISA Extreme
Computing Initiative (www.deisa.org).


\begin{thebibliography}{10}

\bibitem{etmcfollow}
ETMC,
\newblock in preparation  (2007).

\bibitem{Feldmann:1999uf}
T.~Feldmann,
\newblock Int. J. Mod. Phys. {\bf A15}, 159 (2000), [hep-ph/9907491].

\bibitem{McNeile:2000hf}
UKQCD, C.~McNeile and C.~Michael,
\newblock Phys. Lett. {\bf B491}, 123 (2000), [hep-lat/0006020].

\bibitem{Weisz:1982zw}
P.~Weisz,
\newblock Nucl. Phys. {\bf B212}, 1 (1983).

\bibitem{Frezzotti:2000nk}
ALPHA, R.~Frezzotti, P.~A. Grassi, S.~Sint and P.~Weisz,
\newblock JHEP {\bf 08}, 058 (2001), [hep-lat/0101001].

\bibitem{Frezzotti:2003ni}
R.~Frezzotti and G.~C. Rossi,
\newblock JHEP {\bf 08}, 007 (2004), [hep-lat/0306014].

\bibitem{Boucaud:2007uk}
ETMC, P.~Boucaud {\em et~al.},
\newblock Phys. Lett. {\bf B650}, 304 (2007), [hep-lat/0701012].

\bibitem{Jansen:2003ir}
\xlf, K.~Jansen, A.~Shindler, C.~Urbach and I.~Wetzorke,
\newblock Phys. Lett. {\bf B586}, 432 (2004), [hep-lat/0312013].

\bibitem{Jansen:2005kk}
\xlf, K.~Jansen, M.~Papinutto, A.~Shindler, C.~Urbach and I.~Wetzorke,
\newblock JHEP {\bf 09}, 071 (2005), [hep-lat/0507010].

\bibitem{Urbach:2007rt}
ETMC, C.~Urbach,
\newblock PoS {\bf LAT2007}, 022 (2006), [arXiv:0710.1517].

\bibitem{Dimopoulos:2007qy}
ETMC, P.~Dimopoulos, R.~Frezzotti, G.~Herdoiza, C.~Urbach and U.~Wenger,
\newblock PoS {\bf LAT2007}, 102 (2007), [arXiv:0710.2498].

\bibitem{Blossier:2007vv}
ETMC, B.~Blossier {\em et~al.},
\newblock JHEP {\bf 04}, 020 (2008), [arXiv:0709.4574].

\bibitem{Alexandrou:2008tn}
ETMC, C.~Alexandrou {\em et~al.},
\newblock arXiv:0803.3190 [hep-lat].

\bibitem{Jansen:2005cg}
\xlf, K.~Jansen {\em et~al.},
\newblock Phys. Lett. {\bf B624}, 334 (2005), [hep-lat/0507032].

\bibitem{Allton:2004qq}
UKQCD, C.~R. Allton {\em et~al.},
\newblock Phys. Rev. {\bf D70}, 014501 (2004), [hep-lat/0403007].

\bibitem{Shindler:2007vp}
A.~Shindler,
\newblock arXiv:0707.4093 [hep-lat].

\bibitem{Michael:2007vn}
ETMC, C.~Michael and C.~Urbach,
\newblock PoS {\bf LAT2007}, 122 (2007), [arXiv:0709.4564].

\bibitem{Boucaud:2008xu}
ETMC, P.~Boucaud {\em et~al.},
\newblock arXiv:0803.0224 [hep-lat].

\bibitem{Foster:1998vw}
UKQCD, M.~Foster and C.~Michael,
\newblock Phys. Rev. {\bf D59}, 074503 (1999), [hep-lat/9810021].

\bibitem{McNeile:2006bz}
UKQCD, C.~McNeile and C.~Michael,
\newblock Phys. Rev. {\bf D73}, 074506 (2006), [hep-lat/0603007].

\bibitem{Lacock:1994qx}
UKQCD, P.~Lacock, A.~McKerrell, C.~Michael, I.~M. Stopher and P.~W. Stephenson,
\newblock Phys. Rev. {\bf D51}, 6403 (1995), [hep-lat/9412079].

\bibitem{McNeile:2000xx}
UKQCD, C.~McNeile and C.~Michael,
\newblock Phys. Rev. {\bf D63}, 114503 (2001), [hep-lat/0010019].

\bibitem{Gregory:2007ev}
E.~B. Gregory, A.~C. Irving, C.~M. Richards and C.~McNeile,
\newblock Phys. Rev. {\bf D77}, 065019 (2008), [arXiv:0709.4224].

\bibitem{Neff:2001zr}
H.~Neff, N.~Eicker, T.~Lippert, J.~W. Negele and K.~Schilling,
\newblock Phys. Rev. {\bf D64}, 114509 (2001), [hep-lat/0106016].

\bibitem{DeTar:2007ni}
C.~DeTar and L.~Levkova,
\newblock PoS {\bf LAT2007}, 116 (2007), [arXiv:0710.1322 [hep-lat]].

\bibitem{Michael:1982gb}
C.~Michael and I.~Teasdale,
\newblock Nucl. Phys. {\bf B215}, 433 (1983).

\bibitem{Lesk:2002gd}
CP-PACS, V.~I. Lesk {\em et~al.},
\newblock Phys. Rev. {\bf D67}, 074503 (2003), [hep-lat/0211040].

\bibitem{Hashimoto:2008xg}
K.~Hashimoto and T.~Izubuchi,
\newblock arXiv:0803.0186 [hep-lat].

\bibitem{Leutwyler:1992yt}
H.~Leutwyler and A.~Smilga,
\newblock Phys. Rev. {\bf D46}, 5607 (1992).

\bibitem{Edwards:2004sx}
SciDAC, R.~G. Edwards and B.~Joo,
\newblock Nucl. Phys. Proc. Suppl. {\bf 140}, 832 (2005), [hep-lat/0409003].

\bibitem{BAGEL}
P.~Boyle,
\newblock http://www.ph.ed.ac.uk/\~{ }paboyle/bagel/Bagel.html.

\end{thebibliography}

\end{document}